\documentclass[12pt]{article}
\usepackage{amsfonts}
\usepackage{amsmath}

\input{tcilatex}

\begin{document}

\title{Quantized gauge-affine gravity\\
in the superfiber bundle approach}
\author{A. Meziane\thanks{%
E-mail: meziane@univ-oran.dz} \ and \ M. Tahiri \\
Laboratoire de Physique Th\'{e}orique\\
Universit\'{e} d'Oran Es-senia, 31100 Oran, \ Algeria\\
}
\maketitle

\begin{abstract}
The quantization of gauge-affine gravity within the superfiber bundle
formalism is proposed. By introducing an even pseudotensorial 1-superform
over a principal superfiber bundle with superconnection, \ we obtain the
geometrical Becchi-Rouet-Stora-Tyutin (BRST) and anti-BRST transformations
of the fields occurring in such a theory. Reducing the four-dimensional
general affine group double-covering $\overline{GA}(4,%
\mathbb{R}
)$ to the Poincar\'{e} group double-covering $\overline{ISO}(1,3)$ we also
find the BRST and \ anti-BRST transformations of the fields present in
Einstein's \ gravity. \ Furthermore, we give a prescription leading to the
construction of both BRST-invariant gauge-fixing action for gauge-affine
gravity and Einstein's gravity.


PACS \ numbers\textit{: }04.50.+h

DOI: 10.1103/PhysRevD.71.104033
\end{abstract}

\section{\protect\bigskip Introduction}

One of the most outstanding problems in modern theoretical physics is to
construct a consistent\ theory of quantum Einstein gravity. Several models
have been proposed (for a review see e.g. Ref. \cite{mielke4} and references
therein), however none of these models, both \ renormalizable and unitary,
has been found. This is basically due to the dimensionful nature of the
gravitational coupling constant \cite{heinsb2} which destroys the
predictivity of quantum Einstein gravity, i.e. it is impossible to have a
renormalizable theory.

On the other hand, a serious progress has been achieved by Ne'eman and \v{S}%
ija\v{c}ki \cite{neeman1} for solving this problem. They proposed a model
for quantum gravity that reproduces Einstein's gravity at low energy with a
fair possibility to be renormalizable and unitary. However, it has been
proved\ by Stelle in \cite{Ste} that a theory containing Einstein's action
with term quadratic in the appropriate curvatures was renormalisable, but
violated unitarity. The failure of unitarity in that model arises through
the Riemannian condition which relates the connection to the metric (i.e.
torsion free and metric compatible). To avoid this difficulty one has
attempted to consider spacetime with torsion and thus guaranteeing the
independence of metric and connection fields. In this context, the Poincar%
\'{e} gauge theory (PGT) has been developed as a gravitational theory based
on the double-covering $\overline{ISO}(1,3)=SL(2,%
\mathbb{C}
)\otimes 
\mathbb{R}
^{4}$ of the Poincar\'{e} group $ISO(1,3)=SO(1,3)\otimes 
\mathbb{R}
^{4}$ \cite{Hehl1,Hehl2}. We note that the connection is not an independent
variable, since the metricity condition is also preserved in this model \cite%
{Hehl1,Hehl2,hehl112}. However, it has been confirmed that no PGT model can
be renormalizable if one imposes unitarity \cite{kuhfuss}.

Another possibility for doing away with the Riemannian condition consists to
have gravitational gauge model in which the Poincar\'{e} group acting on the
local frames is extended to a larger gauge group for frames, namely$\ 
\overline{GA}(4,%
\mathbb{R}
).$ The resulting gravitational model is a metric-affine gauge theory of
gravity (MAG) which has been suggested in Ref. \cite{neeman1}. The theory
has a metric-affine spacetime with torsion and nonmetricity and incorporates
gravitational models like Einstein's\ gravity. The model is based on gauging
the four-dimensional general affine group $GA(4,%
\mathbb{R}
)=GL(4,%
\mathbb{R}
)\otimes 
\mathbb{R}
^{4}$ \cite{hehl3,hehl4,hehl5}$,$ or its double-covering $\overline{GA}(4,%
\mathbb{R}
)=\overline{GL}(4,%
\mathbb{R}
)\otimes 
\mathbb{R}
^{4}$ \cite{mielke4,mielke44,hehl6}. The existence of a double-covering$%
\overline{\text{ }GL}(4,%
\mathbb{R}
)$ of the general linear group $GL(4,%
\mathbb{R}
)$ has been realized in Ref. \cite{neeman12}. Here, the spinorial
double-covering exists only in infinite matrix representations and the
corresponding infinite-component fields, the so-called manifields \cite%
{neeman12,neeman2}. The renormalizability of MAG model has been proved \cite%
{neeman4,Lee}, but unitarity has not been properly checked to date.

Recently, in \cite{Mielke11} the algebraic structure of
Becchi-Rouet-Stora-Tyutin (BRST) transformations \cite{Bechi} of a
metric-affine gauge gravity based on the Hamiltonian formalism has been
analyzed. This approach leads to the same BRST transformations obtained in 
\cite{neeman22,gron2} in the context of the Batalin and Vilkovisky\
formalism \cite{batalin22}. Here, the authors generalize the work developed
by Okubo \cite{Okubo}\ where a new type of BRST operator has been
constructed only for spacetimes with teleparallelism. They follow the rather
\ transparent exposition of van Holten \cite{holten2} which departs from the
Hamiltonian formalism and replaces the Lagrange multipliers for the first
class constraints by ghost operators.

Moreover, BRST transformations equivalent to those given in \cite%
{Mielke11,neeman22,gron2}\ can also be obtained geometrically. Indeed, as
shown in Ref. \cite{mez}, we have used a superspace formalism to determine
geometrically the BRST and anti-BRST algebra for gauge-affine gravity. Our
method was based on the introduction of $\overline{GA}(4,%
\mathbb{R}
)$-superconnection over a $(4,2)$-dimensional superspace obtained by
extending a metric-affine space with two anticommuting coordinates. This
superconnection represents the gauge fields and their associated ghost and
antighost fields occurring in gauge-affine gravity. In particular, the
introduction of the coordinate ghost and antighost fields leads to the
construction of a basis, where the local expression of the superconnection
becomes more natural. By using this basis, we have determined the BRST and
anti-BRST transformations from the structure equations by imposing
horizontality conditions on the supercurvature.

In the present paper, we discuss the quantization of gauge-affine gravity
theory by using the superfiber bundle formalism, see e.g. Ref. \cite{mez2}
and references therein, in analogy to what is realized for the case of
super-Yang-Mills theory \cite{tah1, tah12} and the four-dimensional
non-Abelian topological antisymmetric tensor gauge theory, the so-called BF
theory \cite{tah2}. In section II, we show how the various fields of\
gauge-affine gravity and their geometrical BRST and anti-BRST
transformations can be determined via a principal superfiber bundle endowed
with a superconnection and an even pseudotensorial 1-superform in the
adjoint representation. Reducing the four-dimensional general affine group
double-covering to the Poincar\'{e} group double-covering we also find the
BRST and anti-BRST transformations of the fields present in quantum Einstein
gravity. The obtained geometrical BRST and anti-BRST transformations are
nilpotent. In section III we first build a gauge-fixing superaction for
gauge-affine gravity as a natural generalization of the one corresponding to
the usual Yang-Mills theory. Then, a gauge-fixing action is obtained as the
lowest component of the gauge-fixing superaction. However, we also work in
the same spirit of construction as in \cite{tah1, tah12,tah2} for building
the gauge-fixing action for quantum Einstein gravity. Section IV is devoted
to concluding remarks.

\section{Geometrical BRST and anti-BRST algebra}

Let $P$ $(M,G_{s})$ be a principal superfiber bundle with superconnection $%
\phi $. The base space $M$ is the four-dimensional metric-affine spacetime
and the structure group $G_{s}$ is the direct product of the general linear
group double-covering\ $\overline{GL}(4,%
\mathbb{R}
)$ with the general affine group double-covering $\overline{GA}(4,%
\mathbb{R}
)$ as well the two-dimensional old translation group $S^{0,2}$.\ We consider 
$P$ as being globally trivial with respect to $S^{0,2}$. This will be
related \ to the fact that the BRST and anti-BRST transformations are
defined globally.

\QTP{Body Math}
The Lie superalgebra $g_{s}$ of the structural Lie supergroup $G_{s}$ is
given by

\QTP{Body Math}
\begin{equation}
g_{s}=\overline{gl}(4,%
\mathbb{R}
)\oplus \overline{ga}(4,%
\mathbb{R}
)\oplus s^{0,2}.  \label{eq1}
\end{equation}

Let $(T_{\sigma }^{\rho })_{\left\{ \rho ,\sigma =1,..,4\right\} },$ $%
(T_{b}^{a})_{\left\{ a,b=1,..,4\right\} },$ ($P_{b})_{\left\{
b=1,..,4\right\} },$\ and $(F_{\alpha })_{\left\{ \alpha =1,2\right\} }$ be
the generators of $\overline{GL}(4,%
\mathbb{R}
),$ $\overline{GA}(4,%
\mathbb{R}
)$\ and $S^{0,2},$ respectively. They satisfy the following commutation
relations

\begin{center}
\begin{eqnarray}
\left[ T_{\varepsilon }^{\tau },T_{\sigma }^{\rho }\right] &=&(\delta
_{\sigma }^{\tau }\delta _{\mu }^{\rho }\delta _{\varepsilon }^{\nu }-\delta
_{\varepsilon }^{\rho }\delta _{\sigma }^{\nu }\delta _{\mu }^{\tau })T_{\nu
}^{\mu },  \notag \\
\left[ T_{b}^{a},T_{d}^{c}\right] &=&(\delta _{d}^{a}\delta _{e}^{c}\delta
_{b}^{f}-\delta _{b}^{c}\delta _{d}^{f}\delta _{e}^{a})T_{f}^{e},  \notag \\
\left[ T_{b}^{a},P_{c}\right] &=&\delta _{c}^{a}\delta _{b}^{d}P_{d},
\label{eq2} \\
\left[ P_{a},P_{b}\right] &=&\left[ T_{b}^{a},T_{\sigma }^{\rho }\right] =%
\left[ P_{a},F_{\alpha }\right] =0,  \notag \\
\left[ T_{\sigma }^{\rho },P_{b}\right] &=&\left[ F_{\alpha },T_{\sigma
}^{\rho }\right] =\left[ T_{b}^{a},F_{\alpha }\right] =\left[ F_{\alpha
},F_{\beta }\right] =0.  \notag
\end{eqnarray}
\end{center}

Let $\Omega $ be an even 2-superform associated to the superconnection $\phi 
$ and $\vartheta $ an even pseudotensorial 1-superform of the type $%
(ad,g_{s})${\footnotesize \footnote{{\footnotesize Here $(ad)$ means adjoint
representation.}}}${\footnotesize .}$ The introduction in $P(M,G_{s})$
besides the usual \ superconnection $\phi $\ an even 1-superform $\vartheta $
will be related, as we will see later, to the fact that the imposed
constraints on the supercurvature $\Omega $ can be obtained by the fact that
the covariant differentiation of a pseudotensorial 1-superform $\vartheta $\
is tensorial.

In order to realize supercurvature constraints, we need to introduce an even
1-superform generalized superconnection $\lambda $ such that

\begin{equation}
\lambda =\phi -\vartheta .  \label{eq3}
\end{equation}%
At this point, let us mention that the introduction of the generalized
superconnection is related, on the one hand, to the fact that the
double-covering group $\overline{Diff}(4,%
\mathbb{R}
)$ of the group of general coordinate transformations (GCT) (i.e., the group
of diffeomorphisms) is realized through the direct product of the general
linear group double-covering $\overline{GL}(4,%
\mathbb{R}
)$ with the translation group $%
\mathbb{R}
^{4}$ as well a simply connected Lie subgroup \cite{neeman2,cant2,siji} and,
on the other hand, as we will see later, to the fact that the gauge-fixing
action for quantum Einstein gravity can be deduced from gauge-fixing action
for gauge-affine gravity by reducing the linear connection to the symmetric
Levi-Civita connection.

Acting the exterior covariant superdifferential $D$ on $\lambda $ we define
then the generalized supercurvature $\Lambda $ ( even 2-superform) given by

\begin{equation}
\Lambda =D\lambda =\Omega -\Theta ,  \label{eq4}
\end{equation}%
where the associated supercurvature $\Omega $ and $\Theta $ to $\phi $ and $%
\vartheta $ are defined by $\Omega =D\phi $ and $\Theta =D\vartheta ,$
respectively. They satisfy the structure equations

\begin{eqnarray}
\Omega &=&d\phi +\frac{1}{2}\left[ \phi ,\phi \right] ,  \label{eq5} \\
\Theta &=&d\vartheta +\left[ \phi ,\vartheta \right] ,  \label{eq6}
\end{eqnarray}%
where $d$ is the exterior superdifferential and $\left[ ,\right] $\ the
graded Lie bracket.

Let$\ z=(z^{M})=(x^{\mu },\theta ^{\alpha })$ be a local coordinates system
on $P,$\ where $(x^{\mu })_{\mu =1,..,4}$\ are the coordinates of the
metric-affine spacetime $M$ and $($ $\theta ^{\alpha })_{\alpha =1,2}$ \ are
ordinary anticommuting variables. Upon expressing the generalized
superconnection $\lambda $\ and the generalized supercurvature $\Lambda $ as

\begin{eqnarray}
\lambda \ &=&dz^{M}\lambda _{M}=\ dz^{M}(\phi _{M}-\vartheta _{M}),  \notag
\\
\Lambda &=&\frac{1}{2}dz^{N}\wedge dz^{M}\Lambda _{MN}=\frac{1}{2}%
dz^{N}\wedge dz^{M}(\Omega _{MN}-\Theta _{MN}),  \label{eq7}
\end{eqnarray}%
we have 
\begin{subequations}
\begin{eqnarray}
\Omega _{MN} &=&\partial _{M}\phi _{N}-(-1)^{mn}\partial _{N}\phi _{M}+\left[
\phi _{M},\phi _{N}\right] ,  \label{eq8} \\
\Theta _{MN} &=&\partial _{M}\vartheta _{N}-(-1)^{mn}\partial _{N}\vartheta
_{M}-\left[ \phi _{N},\vartheta _{M}\right] +(-1)^{mn}\left[ \phi
_{M},\vartheta _{N}\right] ,  \label{eq9}
\end{eqnarray}%
where $m=\mid z^{M}\mid $ is the Grassmann degree of $z^{M}$. Note that the
Grassmann degrees of the superfield components $\phi _{M},$ $\vartheta _{M},$
$\lambda _{M},$ $\Omega _{MN},$ $\Theta _{MN},$ and $\Lambda _{MN}$ are
given by 
\end{subequations}
\begin{eqnarray*}
&\mid &\phi _{M}\mid =\mid \vartheta _{M}\mid =\mid \lambda _{M}\mid =m, \\
&\mid &\Omega _{MN}\mid =\mid \Theta _{MN}\mid =\mid \Lambda _{MN}\mid =m+n%
\text{ (mod 2)},
\end{eqnarray*}%
since $\phi $, $\vartheta $, $\lambda $, $\Omega $, $\Theta $ and $\Lambda $
are even superforms.

Moreover, the generalized supercurvature is a tensorial 2-superform, in
particular we have $i(X)\Lambda =0,$ where $i$ denotes the contraction of
vectors with forms and $X$\ is a vertical superfield in $P.$ Using the fact
that $\partial _{\alpha }=\partial /\partial \theta ^{\alpha \text{ }}$is
vertical, we obtain the following supercurvature equations 
\begin{subequations}
\begin{eqnarray}
\Lambda _{\alpha \beta } &=&0, \\
\Lambda _{\alpha \mu } &=&0,  \label{eq10}
\end{eqnarray}

i.e. 
\end{subequations}
\begin{subequations}
\begin{eqnarray}
\Omega _{\alpha \beta } &=&\Theta _{\alpha \beta },  \label{122} \\
\Omega _{\alpha \mu } &=&\Theta _{\alpha \mu }.  \label{eq12}
\end{eqnarray}%
Furthermore, the $g_{s}-$valued component superfields $\phi _{M},$ $%
\vartheta _{M},$ $\Omega _{MN}$ and $\Theta _{MN}$\ are given by

\end{subequations}
\begin{eqnarray}
\phi _{M} &=&\phi _{bM}^{a}T_{a}^{b}+\phi _{\nu M}^{\mu }T_{\mu }^{\nu
}+\phi _{M}^{a}P_{a}+\phi _{M}^{\alpha }F_{\alpha },  \notag \\
\vartheta _{M} &=&\vartheta _{bM}^{a}T_{a}^{b}+\vartheta _{\nu M}^{\mu
}T_{\mu }^{\nu }+\vartheta _{M}^{a}P_{a}+\vartheta _{M}^{\alpha }F_{\alpha },
\label{eq14} \\
\Omega _{MN} &=&\Omega _{bMN}^{a}T_{a}^{b}+\Omega _{\nu MN}^{\mu }T_{\mu
}^{\nu }+\Omega _{MN}^{a}P_{a}+\Omega _{MN}^{\alpha }F_{\alpha },  \notag \\
\Theta _{MN} &=&\Theta _{bMN}^{a}T_{a}^{b}+\Theta _{\nu MN}^{\mu }T_{\mu
}^{\nu }+\Theta _{MN}^{a}P_{a}+\Theta _{MN}^{\alpha }F_{\alpha }.  \notag
\end{eqnarray}

According to Eqs (10) and (\ref{eq14})\ we find

\begin{subequations}
\begin{eqnarray}
\Omega _{b\mu \alpha }^{a} &=&\Theta _{b\mu \alpha }^{a},  \label{eq.137} \\
\Omega _{b\alpha \beta }^{a} &=&\Theta _{b\alpha \beta }^{a}, \\
\Omega _{\sigma \mu \alpha }^{\rho } &=&\Theta _{\sigma \mu \alpha }^{\rho },
\label{eq134} \\
\Omega _{\sigma \alpha \beta }^{\rho } &=&\Theta _{\sigma \alpha \beta
}^{\rho },  \label{eq13} \\
\Omega _{\mu \beta }^{a} &=&\Theta _{\mu \beta }^{a},  \label{136} \\
\Omega _{\alpha \beta }^{a} &=&\Theta _{\alpha \beta }^{a}, \\
\Omega _{\mu \beta }^{\alpha } &=&\Theta _{\mu \beta }^{\alpha },
\label{eq.138} \\
\Omega _{\gamma \beta }^{\alpha } &=&\Theta _{\gamma \beta }^{\alpha }.
\label{eq.139}
\end{eqnarray}%
The components $\Lambda _{MN}^{\alpha }$ associated to $F_{\alpha \text{ \ }%
} $give the $S^{0,2}$-generalized supertorsion 
\end{subequations}
\begin{equation}
\Lambda _{MN}^{\alpha }=\Omega _{MN}^{\alpha }-\Theta _{MN}^{\alpha }.
\label{eqtorsion}
\end{equation}%
According to (\ref{eq.138}) and (\ref{eq.139}) suppplemented with the
constraint

\begin{equation*}
\Lambda _{\mu \nu }^{\alpha }=0,
\end{equation*}%
we find then that the $S^{0,2}$-generalized supertorsion vanishes. Moreover,
the potentials $\phi _{M\text{ \ }}^{\alpha }$ being pure gauge, we can then
impose the following supercurvature constraint 
\begin{equation}
\Omega _{MN}^{\alpha }=\partial _{M}\phi _{N}^{\alpha }-(-1)^{mn}\partial
_{N}\phi _{M}^{\alpha }=0,  \label{eq255}
\end{equation}%
and therefore we also have%
\begin{equation}
\Theta _{MN}^{\alpha }=\partial _{M}\vartheta _{N}^{\alpha
}-(-1)^{mn}\partial _{N}\vartheta _{M}^{\alpha }=0.  \label{eq256}
\end{equation}

In addition, we impose that the geometrical structure of the principal
superfiber bundle $P(M,G_{s})$ should incorporate the metric-affine
structure of the spacetime $M$ such that for $\theta ^{\alpha }=0$, the
components of the superfields $\Lambda _{\mu \nu }$ permit us to find the
standard results concerning the torsion and the curvature of metric-affine
spacetime. This allows us to have

\begin{equation}
\Lambda _{\mu \nu }=\Omega _{\mu \nu },  \label{eqstand}
\end{equation}%
and therefore

\begin{equation}
\Theta _{\tau \mu \nu }^{\rho }=\Theta _{b\mu \nu }^{a}=\Theta _{\mu \nu
}^{a}=\Theta _{\mu \nu }^{\alpha }=0.  \label{eqstand2}
\end{equation}

Since the potentials $\phi _{M\text{ \ }}^{\alpha }$being pure gauge, we
consider, hereafter, that the components $\phi _{M}$ are $\overline{gl}(4,%
\mathbb{R}
)\oplus \overline{ga}(4,%
\mathbb{R}
)$-valued superfields and can be written as

\begin{equation}
\phi =dz^{M}\phi _{M}=dx^{\mu }\phi _{\mu }+d\theta ^{\alpha }\eta _{\alpha
},  \label{eq26}
\end{equation}%
where

\begin{subequations}
\begin{eqnarray}
\phi _{\mu } &=&\phi _{b\mu }^{a}T_{a}^{b}+\phi _{\tau \mu }^{\rho }T_{\rho
}^{\tau }+\phi _{\mu }^{a}P_{a},  \label{eq27} \\
\eta _{\alpha } &=&\eta _{b\alpha }^{a}T_{a}^{b}+\eta _{\tau \alpha }^{\rho
}T_{\rho }^{\tau }+\eta _{\alpha }^{a}P_{a}.  \label{eq28}
\end{eqnarray}

Now, in order to derive the BRST structure of gauge-affine gravity it is
necessary to give the geometrical description of the fields present in such
theory. To this purpose, we assign to the anticommuting coordinates $\theta
^{1}$ and $\theta ^{2\text{ }}$the ghost numbers $(-1)$ and $(+1),$
respectively, and ghost number zero for an even quantity: either a
coordinate, a superform or a generator. These rules permit us to determine
the ghost numbers of the superfields $(\phi _{\tau \mu }^{\rho },$ $\phi
_{b\mu }^{a},$ $\phi _{\mu }^{a},$ $\eta _{b1}^{a},$ $\eta _{b2}^{a})$ which
are given by $(0,0,0,1,-1).$ So, the lowest components $\phi _{\tau \mu \mid
}^{\rho },$ $\phi _{b\mu \mid }^{a},$ $\phi _{\mu \mid }^{a},$ $\eta
_{b1\mid }^{a}$, and $\ \eta _{b2\mid }^{a}$ can be identified with the
linear connection $\Gamma _{\tau \mu }^{\rho }$, the affine connection $%
\omega _{b\mu }^{a}$, the vierbein $e_{\mu }^{a}$, the $\overline{GA}(4,%
\mathbb{R}
)$ ghost $c_{b}^{a}$ and its antighost $\overline{c}_{b}^{a},$
respectively.\ The symbol $``\mid $\textquotedblright\ indicates that the
superfield is evaluated at $\theta ^{\alpha }=0.$ Moreover, we\ introduce
the coordinate (diffeomorphism) ghost and antighost superfields $\eta
_{\alpha }^{\mu }$ by the following replacement

\end{subequations}
\begin{subequations}
\begin{eqnarray}
\eta _{\tau \alpha }^{\mu } &=&\partial _{\tau }\eta _{\alpha }^{\mu }+\phi
_{\tau \rho }^{\mu }\eta _{\alpha }^{\rho }  \label{eq29} \\
\eta _{\alpha }^{a} &=&0.  \label{eq30}
\end{eqnarray}%
This permits us, on the one hand, to introduce the coordinate ghost $c^{\mu
}=\eta _{1\mid }^{\mu }$ and its antighost $\overline{c}^{\mu }=\eta _{2\mid
}^{\mu }$\ and, on the other hand, to justify the introduction of general
linear group double-covering $\overline{GL}(4,%
\mathbb{R}
)$ with generators $(T_{\sigma }^{\rho })$ in the structure group $G_{s}$\
of the principal superfiber bundle $P$ $(M,G_{s}).$ Furthermore, knowing the
expression of the components of $\Theta $\ we can determine the components
of $\vartheta $ by using the relation (\ref{eq9}). Some components of $%
\Theta $\ have been determined from equations (\ref{eq256})\ and (\ref%
{eqstand2}). As trivial solutions we have

\end{subequations}
\begin{equation}
\vartheta _{\beta }^{\alpha }=\vartheta _{\mu }^{\alpha }=\vartheta _{\mu
}^{a}=\vartheta _{b\mu }^{a}=\vartheta _{\tau \mu }^{\rho }=0.  \label{eq39}
\end{equation}%
The determination of the others components of $\vartheta $ such $\vartheta
_{b\alpha }^{a},$ $\vartheta _{\rho \alpha }^{\tau },$ and $\vartheta
_{\beta }^{a}$\ can be easily obtained from the remaining $\Theta $\
components. The latters, after straightforward calculations, acquire the
form 
\begin{subequations}
\begin{eqnarray}
\Theta _{b\mu \alpha }^{a} &=&\eta _{\alpha }^{\rho }\Omega _{b\rho \mu
}^{a}+D_{\mu }\{\phi _{b\rho }^{a}\eta _{\alpha }^{\rho }\},  \label{eq405}
\\
\Theta _{\tau \alpha \beta }^{\rho } &=&-\frac{1}{2}\left[ \eta _{\alpha
}^{\sigma },\eta _{\beta }^{\nu }\right] \Omega _{\tau \sigma \nu }^{\rho },
\label{eq402} \\
\Theta _{b\alpha \beta }^{a} &=&\left\{ 
\begin{array}{c}
-2\eta _{\alpha }^{\rho }\partial _{\rho }\eta _{b\beta }^{a}\text{ if }%
\alpha =\beta , \\ 
{\LARGE 0}\text{ if }\alpha \neq \beta ,%
\end{array}%
\right.  \label{eq403} \\
\Theta _{\mu \alpha }^{a} &=&\eta _{\alpha }^{\rho }\Omega _{\rho \mu
}^{a}+D_{\mu }\{\eta _{\alpha }^{\rho }\phi _{\rho }^{a}\},  \label{eq404} \\
\Theta _{\alpha \beta }^{a} &=&0,  \label{eq406}
\end{eqnarray}%
where $D_{\mu }=\partial _{\mu }+\left[ \phi _{\mu },\right] $ is the $%
\overline{ga}(4,%
\mathbb{R}
)$-valued covariant superderivative. It is worth noting that the anholonomic
and holonomic components of the superconnection $\phi _{\mu \nu }^{\sigma }$
and $\phi _{b\mu }^{a}$ are related by the supervierbein $\phi _{\mu }^{a}$\
as follows 
\end{subequations}
\begin{equation}
\phi _{\mu \nu }^{\sigma }=\phi _{a}^{\sigma }(\partial _{\nu }\phi _{\mu
}^{a}\ +\phi _{\mu }^{b}\phi _{b\nu }^{a}).  \label{holonomic}
\end{equation}%
Therefore, the components$\ \Theta _{\tau \mu \alpha }^{\rho }$ can be
derived from the components $\ \Theta _{b\mu \alpha }^{a}$ and $\ \Theta
_{\mu \alpha }^{a}$ as follows

\begin{equation}
\ \Theta _{\tau \mu \alpha }^{\rho }=\eta _{\alpha }^{\nu }\Omega _{\tau \mu
\nu }^{\rho }.  \label{eq4066}
\end{equation}

However, the operational representation for an infinitesimal $S^{0,2}$%
-motion in $P$ is given by

\begin{equation}
r(\theta ^{\alpha })=1+\theta ^{\alpha }Q_{\alpha },  \label{eq31}
\end{equation}%
where $(Q_{\alpha })_{\alpha =1,2}$ are the differential operators
representing the $S^{0,2}$-generators $(F_{\alpha }).$ According to the fact
that the superconnection is a pseudotensorial 1-superform in the adjoint
representation, we have

\begin{equation}
\phi _{M}^{A}(x^{\mu },\zeta ^{\alpha }+\theta ^{\alpha })=r(\theta ^{\alpha
})\phi _{M}^{A}(x^{\mu },\zeta ^{\alpha })r^{-1}(\theta ^{\alpha }).
\label{eq32}
\end{equation}%
It is straightforward to compute (\ref{eq32}), and we find

\begin{equation}
\phi _{M}^{A}(x^{\mu },\zeta ^{\alpha }+\theta ^{\alpha })=\phi
_{M}^{A}(x^{\mu },\zeta ^{\alpha })+\theta ^{\alpha }\left[ Q_{\alpha },\phi
_{M}^{A}(x^{\mu },\zeta ^{\alpha })\right] +\frac{1}{2}\theta ^{\alpha
}\theta ^{\beta }\left[ Q_{\beta ,}\left[ Q_{\alpha },\phi _{M}^{A}(x^{\mu
},\zeta ^{\alpha })\right] \right] .  \label{eq33}
\end{equation}%
By expanding $\phi _{M}^{A}(x^{\mu },\theta ^{\alpha })$ in power series of $%
\theta ^{\alpha },$ we have

\begin{subequations}
\begin{eqnarray}
\phi _{\tau \mu }^{\sigma } &=&\Gamma _{\tau \mu }^{\sigma }+\theta ^{\alpha
}A_{\tau \mu \alpha }^{\sigma }+\frac{1}{2}\theta ^{\alpha }\theta ^{\beta
}B_{\tau \mu \beta \alpha }^{\sigma },  \label{eq375} \\
\phi _{b\mu }^{a} &=&\omega _{b\mu }^{a}+\theta ^{\alpha }M_{b\mu \alpha
}^{a}+\frac{1}{2}\theta ^{\alpha }\theta ^{\beta }N_{b\mu \beta \alpha }^{a},
\label{eq372} \\
\phi _{\mu }^{a} &=&e_{\mu }^{a}+\theta ^{\alpha }K_{\mu \alpha }^{a}+\frac{1%
}{2}\theta ^{\alpha }\theta ^{\beta }L_{\mu \beta \alpha }^{a},  \label{eq37}
\\
\eta _{b\delta }^{a} &=&c_{b\delta }^{a}+\theta ^{\alpha }R_{b\delta \alpha
}^{a}+\frac{1}{2}\theta ^{\alpha }\theta ^{\beta }S_{b\delta \beta \alpha
}^{a},  \label{eq374} \\
\eta _{\delta }^{\mu } &=&c_{\delta }^{\mu }+\theta ^{\alpha }V_{\delta
\alpha }^{\mu }+\frac{1}{2}\theta ^{\alpha }\theta ^{\beta }W_{\delta \beta
\alpha }^{\mu },  \label{eq376}
\end{eqnarray}%
where $B_{\tau \mu \beta \alpha }^{\sigma },$ $N_{b\mu \beta \alpha }^{a},$ $%
L_{\mu \beta \alpha }^{a},\ S_{b\delta \beta \alpha }^{a}$ and $W_{\delta
\beta \alpha }^{\mu }$ are antisymmetric with respect to the indices $\alpha 
$\ and $\beta .$ Evaluating (\ref{eq33}) at$\ \zeta ^{\alpha }=0$ and in
view of Eq. (28), we obtain

\end{subequations}
\begin{subequations}
\begin{eqnarray}
A_{\tau \mu \alpha }^{\sigma } &=&\left[ Q_{\alpha },\Gamma _{\tau \mu
}^{\sigma }\right] =\partial _{\alpha }\phi _{\tau \mu \mid }^{\sigma },
\label{eq347} \\
M_{b\mu \alpha }^{a} &=&\left[ Q_{\alpha },\omega _{b\mu }^{a}\right]
=\partial _{\alpha }\phi _{b\mu \mid }^{a},  \label{eq346} \\
K_{\mu \alpha }^{a} &=&\left[ Q_{\alpha },e_{\mu }^{a}\right] =\partial
_{\alpha }\phi _{\mu \mid }^{a},  \label{eq343} \\
R_{b\delta \alpha }^{a} &=&\left[ Q_{\alpha },c_{b\delta }^{a}\right]
=\partial _{\alpha }\eta _{b\delta \mid }^{a},  \label{eq348} \\
V_{\delta \alpha }^{\mu } &=&\left[ Q_{\alpha },c_{\delta }^{\mu }\right]
=\partial _{\alpha }\eta _{\delta \mid }^{\mu }.  \label{eq349}
\end{eqnarray}%
We also obtain similar relations for the other field components

\end{subequations}
\begin{subequations}
\begin{eqnarray}
B_{\tau \mu \beta \alpha }^{\sigma } &=&\left[ Q_{\beta ,}\left[ Q_{\alpha
},\Gamma _{\tau \mu }^{\sigma }\right] \right] =\partial _{\beta }\partial
_{\alpha }\phi _{\tau \mu \mid }^{\sigma },  \label{eq3441} \\
N_{b\mu \beta \alpha }^{a} &=&\left[ Q_{\beta ,}\left[ Q_{\alpha },\omega
_{b\mu }^{a}\right] \right] =\partial _{\beta }\partial _{\alpha }\phi
_{b\mu \mid }^{a},  \label{eq3442} \\
L_{\mu \beta \alpha }^{a} &=&\left[ Q_{\beta ,}\left[ Q_{\alpha },e_{\mu
}^{a}\right] \right] =\partial _{\beta }\partial _{\alpha }\phi _{\mu \mid
}^{a},  \label{eq34} \\
S_{b\delta \beta \alpha }^{a} &=&\left[ Q_{\beta ,}\left[ Q_{\alpha
},c_{b\delta }^{a}\right] \right] =\partial _{\beta }\partial _{\alpha }\eta
_{b\delta \mid }^{a},  \label{eq3444} \\
W_{\delta \beta \alpha }^{\mu } &=&\left[ Q_{\beta ,}\left[ Q_{\alpha
},c_{\delta }^{\mu }\right] \right] =\partial _{\beta }\partial _{\alpha
}\eta _{\delta \mid }^{\mu }.  \label{eq3445}
\end{eqnarray}%
In analogy with the Yang-Mills case \cite{tah12}, we remark that the
operators $Q_{1}$ and $Q_{2\text{ }}$represent the BRST and anti-BRST
operators $Q$ and $\overline{Q}$, respectively.

Evaluating (\ref{eq8}) at $\theta ^{\alpha }=0$ and using (12), (20), (22),
(28) and (29) we obtain the following geometrical BRST transformations

\end{subequations}
\begin{eqnarray}
\left[ Q,\Gamma _{\tau \mu }^{\sigma }\right] &=&\Gamma _{\tau \mu }^{\rho
}\partial _{\rho }c^{\sigma }-\partial _{\mu }\partial _{\tau }c^{\sigma
}-\Gamma _{\tau \rho }^{\sigma }\partial _{\mu }c^{\rho }-\Gamma _{\rho \mu
}^{\sigma }\partial _{\tau }c^{\rho }-c^{\rho }\partial _{\rho }\Gamma
_{\tau \mu }^{\sigma },  \notag \\
\left[ Q,\omega _{b\mu }^{a}\right] &=&\partial _{\mu
}c_{b}^{a}+c_{b}^{d}\omega _{d\mu }^{a}-c_{d}^{a}\omega _{b\mu }^{d}+\omega
_{b\sigma }^{a}\partial _{\mu }c^{\sigma }+c^{\rho }\partial _{\rho }\omega
_{b\mu }^{a},  \notag \\
\left[ Q,e_{\mu }^{a}\right] &=&e_{\sigma }^{a}\partial _{\mu }c^{\sigma
}+c^{\rho }\partial _{\rho }e_{\mu }^{a}-e_{\mu }^{b}c_{b}^{a},  \notag \\
\left[ Q,c_{b}^{a}\right] &=&c^{\rho }\partial _{\rho
}c_{b}^{a}-c_{f}^{a}c_{b}^{f},  \notag \\
\left[ Q,c^{\sigma }\right] &=&c^{\rho }\partial _{\rho }c^{\sigma },
\label{eqbrst} \\
\left[ Q,\overline{c}_{b}^{a}\right] &=&B_{b}^{a},  \notag \\
\left[ Q,\overline{c}^{\sigma }\right] &=&B^{\sigma },  \notag \\
\left[ Q,B_{b}^{a}\right] &=&0,  \notag \\
\left[ Q,B^{\sigma }\right] &=&0,  \notag
\end{eqnarray}%
and also the geometrical anti-BRST transformations, which can be derived
from (\ref{eqbrst}) by the following rules: $X\longrightarrow X,$ if $X$= $%
\Gamma _{\tau \mu }^{\sigma },$ $\omega _{b\mu }^{a},$ $e_{\mu }^{a};$ $\
X\longrightarrow \overline{X},$ if $X$= $Q,$ $c^{\mu },$ $c_{b}^{a},$ $%
B^{\mu },$ $B_{b}^{a}$ and $X=\overline{\overline{X}},$ where

\begin{eqnarray}
B^{\mu }+\overline{B}^{\mu } &=&c^{\rho }\partial _{\rho }\overline{c}^{\mu
}+\overline{c}^{\rho }\partial _{\rho }c^{\mu },  \notag \\
B_{b}^{a}+\overline{B}_{b}^{a} &=&c^{\rho }\partial _{\rho }\overline{c}%
_{b}^{a}+\overline{c}^{\rho }\partial _{\rho }c_{b}^{a}-c_{d}^{a}\overline{c}%
_{b}^{d}-\overline{c}_{d}^{a}c_{b}^{d}.  \label{eqbrst2}
\end{eqnarray}

Let us note that the obtained BRST and anti-BRST transformations are
nilpotent, i.e. 
\begin{equation}
Q^{2}=\overline{Q}^{2}=\left[ Q,\overline{Q}\right] =0.  \label{eqbrst33}
\end{equation}

Now, we apply the same geometrical framework to find the BRST and anti-BRST
transformations of the fields occurring in quantum Einstein gravity. To this
end, we must reduce the general affine group double-covering $\overline{GA}%
(4,%
\mathbb{R}
)$ to the Poincar\'{e} double-covering$\overline{\text{ }ISO}(1,3)$. The
BRST transformations of the fields associated to Poincar\'{e}
double-covering have already be given in \cite{nakanishi}. Reducing $%
\overline{GA}(4,%
\mathbb{R}
)$ to $\overline{ISO}(1,3)$ leads us to keep from (12) only

\begin{equation}
\Omega _{\sigma \alpha \beta }^{\rho }=\Theta _{\sigma \alpha \beta }^{\rho
}.  \label{eqmetri10}
\end{equation}

On the other hand, Einstein's theory is Riemannian, i.e. it precludes the
propagation of\ either torsion or nonmetricity. Only the coordinate metric
field $g_{\mu \nu \text{\ }}$propagates. Here the coordinate metric field $%
g_{\mu \nu \text{\ }}$is related to the Minkowski metric $\eta _{ab\text{ }}$%
through the vierbein $e_{\mu }^{a}$ as follows

\begin{equation}
g_{\mu \nu \text{\ }}=\eta _{ab\text{ }}e_{\mu }^{a}e_{\nu }^{b},
\label{metric}
\end{equation}%
and can be written, in view of (\ref{eq37}), as a lowest component of a
superfield $G_{\mu \nu }$ which can be put in the form%
\begin{equation*}
G_{\mu \nu }=g_{\mu \nu \text{\ }}+\theta ^{\alpha }H_{\mu \nu \alpha }+%
\frac{1}{2}\theta ^{\alpha }\theta ^{\beta }E_{\mu \nu \beta \alpha },
\end{equation*}%
where $H_{\mu \nu \alpha }$ and $E_{\mu \nu \beta \alpha }$ follow from (\ref%
{eq37}). This remark permits us to find the BRST transformation of the
coordinate metric field $g_{\mu \nu \text{\ }}$through the BRST
transformation of the vierbein $e_{\mu }^{a}$

\begin{equation*}
\left[ Q,g_{\mu \nu \text{\ }}\right] =\eta _{ab\text{ }}\left[ Q,e_{\mu
}^{a}\right] e_{\nu }^{b}+\eta _{ab\text{ }}e_{\mu }^{a}\left[ Q,e_{\nu }^{b}%
\right] .
\end{equation*}%
The latter becomes%
\begin{equation}
\left[ Q,g_{\mu \nu \text{\ }}\right] =g_{\mu \sigma \text{\ }}\partial
_{\nu }c^{\sigma }+c^{\rho }\partial _{\rho }g_{\mu \nu \text{\ }}+g_{\sigma
\nu \text{\ }}\partial _{\mu }c^{\sigma },  \label{metricbrst}
\end{equation}%
by using (\ref{eqbrst}) and the fact that $c_{bd}=-c_{db}.$

Moreover, according to (\ref{eq402}), (\ref{eq376}) and (\ref{eqmetri10}),
we obtain

\begin{eqnarray}
V_{\delta \alpha }^{\mu } &=&c_{\delta }^{\rho }\partial _{\rho }c_{\alpha
}^{\mu },  \notag \\
V_{12}^{\tau }+V_{21}^{\tau } &=&c_{1}^{\rho }\partial _{\rho }c_{2}^{\tau
}+c_{2}^{\rho }\partial _{\rho }c_{1}^{\tau }.  \label{metricbrst2}
\end{eqnarray}

Therefore, making use of (\ref{eq349}) and keeping the same identifications,
we find the following BRST transformations \cite{nakanishi}%
\begin{eqnarray}
\left[ Q,c^{\sigma }\right] &=&c^{\rho }\partial _{\rho }c^{\sigma },  \notag
\\
\left[ Q,\overline{c}^{\sigma }\right] &=&B^{\sigma },  \label{cord2} \\
\left[ Q,B^{\sigma }\right] &=&0.  \notag
\end{eqnarray}%
We also obtain the geometrical anti-BRST transformations, which can be
derived from (\ref{metricbrst}) and (\ref{cord2}) by the following mirror
symmetry of the ghost numbers: $X\longrightarrow X,$ if $X\ =g_{\mu \nu 
\text{\ }},\ X\longrightarrow \overline{X},$ if$\ X$= $Q,$ $c^{\mu },$ $%
B^{\mu }$ and $\overline{\overline{X}}\longrightarrow $ $X.$

\section{Gauge-fixing quantum action}

In the present section, we show how to construct a BRST-invariant
gauge-fixing quantum action for gauge-affine gravity as the lowest component
of a gauge-fixing superaction. To this purpose, we\ propose starting with a
gauge-fixing superaction similar to that obtained in the case of
super-Yang-Mills theory as given in \cite{tah1,tah12}

\begin{eqnarray}
S_{sgf} &=&\dint d^{4}xL_{sgf,}  \notag \\
L_{sgf} &=&(\partial _{1}\phi _{2})(\partial ^{\mu }\phi _{\mu })+(\partial
^{\mu }\phi _{2})(\partial _{1}\phi _{\mu })+(\partial _{1}\phi
_{2})(\partial _{1}\phi _{2}).  \label{fixing1}
\end{eqnarray}%
We note first that it is the superconnection $\phi $ which is $\overline{gl}%
(4,%
\mathbb{R}
)\oplus \overline{ga}(4,%
\mathbb{R}
)$-valued and represents the fields occurring in quantized gauge-affine
gravity. This allows us to write the superaction $S_{sgf}$ \ as follows

\begin{equation}
S_{sgf}=S_{sgf}^{t}+S_{sgf}^{d},  \label{fixing2}
\end{equation}%
where $S_{sgf}^{t}$ and$\ S_{sgf}^{d}$ are associated to the tangent and
spacetime indices, respectively.

According to the fact that

\begin{eqnarray*}
\partial _{1}\phi _{b2\mid }^{a} &=&B_{b}^{a}=\left[ Q,\overline{c}_{b}^{a}%
\right] , \\
\partial _{1}\phi _{b\mu \mid }^{a} &=&\left[ Q,\omega _{b\mu }^{a}\right]
\end{eqnarray*}%
and in view of (\ref{fixing1}), we can write the gauge-fixing action
associated to the tangent indices, $S_{gf}^{t}=S_{sgf\mid }^{t},$ in the
following form

\begin{equation}
S_{gf}^{t}=\dint d^{4}x\left\{ B_{b}^{a}\partial ^{\mu }\omega _{a\mu
}^{b}+\partial ^{\mu }\overline{c}_{b}^{a}\left[ Q,\omega _{a\mu }^{b}\right]
+B_{b}^{a}B_{a}^{b}\right\} .  \label{fixing3}
\end{equation}%
Concerning the superaction $S_{sgf}^{d}$ \ we note\ that the antighost$\ 
\overline{c}^{\rho }$ of the general coordinate transformations and the
auxiliary field $B^{\rho }$ are introduced through the relations (\ref{eq29}%
) and (\ref{eq376}).\ Thus, the superaction $S_{sgf}^{d}$ can be obtained
from the prescription (\ref{fixing1}) by substituting the component
superfield $\phi _{\tau 2}^{\rho }$ with the antighost superfield ($\eta
_{2}^{\rho })$ and using the necessary contraction of the components $\phi
_{\tau \mu \text{ }}^{\rho }.$ This gives%
\begin{equation}
S_{sgf}^{d}=\dint d^{4}x{\LARGE (}\partial _{1}\eta _{2}^{\rho }\partial
^{\mu }\phi _{\rho \mu }^{\rho }+\partial ^{\mu }\eta _{2}^{\rho }\partial
_{1}\phi _{\rho \mu }^{\rho }+\partial _{1}\eta _{2}^{\rho }\partial
_{1}\eta _{2}^{\rho }{\LARGE )}.  \label{fixing4}
\end{equation}%
Using the fact that

\begin{equation*}
\partial _{1}\eta _{2\mid }^{\rho }=B^{\rho }=\left[ Q,\overline{c}^{\rho }%
\right]
\end{equation*}%
and%
\begin{equation*}
\partial _{1}\phi _{\rho \mu \mid }^{\rho }=\left[ Q,\Gamma _{\rho \mu
}^{\rho }\right] ,
\end{equation*}%
we obtain%
\begin{equation}
S_{sgf\mid }^{d}=S_{gf}^{d}=\dint d^{4}x{\LARGE (}B^{\rho }\partial ^{\mu
}\Gamma _{\rho \mu }^{\rho }+\partial ^{\mu }\overline{c}^{\rho }\left[
Q,\Gamma _{\rho \mu }^{\rho }\right] +B^{\rho }B_{\rho }{\LARGE )}.
\label{fixing5}
\end{equation}%
Then, it is quite easy to show that the gauge-fixing action,

\begin{equation}
S_{gf}=S_{gf}^{t}+S_{gf}^{d},
\end{equation}%
is invariant with respect to the geometrical BRST transformations. In fact,
\ we have

\begin{eqnarray}
\left[ Q,S_{gf}^{t}\right] &=&\int d^{4}x{\LARGE (}B_{b}^{a}\left[
Q,\partial ^{\mu }\omega _{a\mu }^{b}\right] +\left[ Q,\partial ^{\mu }%
\overline{c}_{b}^{a}\right] \left[ Q,\omega _{a\mu }^{b}\right] {\LARGE )}, 
\notag \\[0.01in]
\left[ Q,S_{gf}^{d}\right] &=&\int d^{4}x{\LARGE (}B^{\rho }\left[
Q,\partial ^{\mu }\Gamma _{\rho \mu }^{\rho }\right] +\left[ Q,\partial
^{\mu }\overline{c}^{\rho }\right] \left[ Q,\Gamma _{\rho \mu }^{\rho }%
\right] {\LARGE )},  \label{fixing67}
\end{eqnarray}%
and using the fact that the geometrical BRST operator $Q$ commutes with the
differential operator we get

\begin{equation}
\left[ Q,S_{gf}\right] =\int d^{4}x\left\{ \partial ^{\mu }{\large (}%
B_{b}^{a}\left[ Q,\omega _{a\mu }^{b}\right] +B^{\rho }\left[ Q,\Gamma
_{\rho \mu }^{\rho }\right] {\Large )}\right\} .
\end{equation}

From this, it follows that the $Q$ \ invariance of $S_{gf}$ \ is guaranteed
modulo a total divergence. So we have constructed the $Q-$invariant
gauge-fixing action for gauge-affine gravity theory \cite{Lee,neeman22}.

Furthermore, it is also interesting to construct the gauge-fixing action for
quantum Einstein gravity in analogy with what is realized in
super-Yang-Mills theory \cite{tah1,tah12}. Let us first remark that the
expression (\ref{fixing3}) corresponds to the gauge-fixing action $%
S_{gf}^{t} $ \ associated to the tangent $\overline{ISO}(1,3)$ group where
we should consider the BRST transformation $\left[ Q,\omega _{a\mu }^{b}%
\right] $\ as in (\ref{eqbrst}), see also Ref. \cite{nakanishi}. To
determine the gauge-fixing action $S_{gf}^{d}$ associated to the
diffeomorphisms group, we proceed as below but we should substitute in (\ref%
{fixing1}) the superfield components of the superconnection by the dynamic
fields occurring in quantum Einstein gravity, namely the antighost $%
\overline{c}^{\mu }$ and the metric $g_{\rho \mu }$ which is introduced by $%
\widetilde{g}_{\rho \mu }=\sqrt{-g}g_{\rho \mu }$, where $g$ is the
determinant of $g_{\rho \mu },$ we obtain

\begin{equation}
S_{gf}^{d}=\dint d^{4}x{\LARGE (}B^{\rho }\partial ^{\mu }\widetilde{g}%
_{\rho \mu }+\partial ^{\mu }\overline{c}^{\rho }\left[ Q,\widetilde{g}%
_{\rho \mu }\right] +B^{\rho }B_{\rho }{\LARGE ),}  \label{fixing 4}
\end{equation}%
where the BRST transformation of the field$\widetilde{\text{ }g}_{\rho \mu }$
is given by%
\begin{equation}
\left[ Q,\widetilde{g}_{\rho \mu }\right] =\partial _{\sigma }(c^{\sigma }%
\widetilde{g}_{\rho \mu })+\widetilde{g}_{\sigma \mu }\partial _{\rho
}c^{\sigma }+\widetilde{g}_{\sigma \rho }\partial _{\mu }c^{\sigma }.
\label{coordbrst}
\end{equation}%
In fact, knowing that $\widetilde{g}_{\rho \mu }=\sqrt{-g}g_{\rho \mu }$,\
we have 
\begin{equation*}
\left[ Q,\widetilde{g}_{\rho \mu }\right] =\left[ Q,\sqrt{-g}\right] g_{\rho
\mu }+\sqrt{-g}\left[ Q,g_{\rho \mu }\right] .
\end{equation*}%
Then by using the fact that

\begin{eqnarray*}
\left[ Q,\sqrt{-g}\right] &=&\frac{-1}{2\sqrt{-g}}\left[ Q,g\right] , \\
\left[ Q,g\right] &=&gg^{\mu \nu }\left[ Q,g_{\mu \nu }\right] =2g\partial
_{\sigma }c^{\sigma }+c^{\sigma }\partial _{\sigma }g,
\end{eqnarray*}%
\bigskip we have%
\begin{equation*}
\left[ Q,\sqrt{-g}\right] =\partial _{\sigma }(c^{\sigma }\sqrt{-g}),
\end{equation*}%
and so we can easily derive the relation (\ref{coordbrst}). Finally\ the
BRST-invariant gauge-fixing action $S_{gf}^{d}$ associated to the
diffeomorphisms group can be written as follows \cite{nakanishi}

\begin{equation}
S_{gf}^{d}=\dint d^{4}x\left\{ B^{\rho }\partial ^{\mu }\widetilde{g}_{\rho
\mu }+\partial ^{\mu }\overline{c}^{\rho }(\partial _{\sigma }(c^{\sigma }%
\widetilde{g}_{\rho \mu })+\widetilde{g}_{\sigma \mu }\partial _{\rho
}c^{\sigma }+\widetilde{g}_{\sigma \rho }\partial _{\mu }c^{\sigma
})+B^{\rho }B_{\rho }\right\} .  \label{fixingcord}
\end{equation}

\section{Conclusion}

In the present paper a geometric formulation of quantized gauge-affine
gravity has been provided using a superfiber bundle formalism with base
space simply the metric-affine spacetime and a structure group the direct
product of the general linear group double-covering\ $\overline{GL}(4,%
\mathbb{R}
)$ with the general affine group double-covering $\overline{GA}(4,%
\mathbb{R}
)$ as well the two-dimensional old translation group $S^{0,2}$. In this
geometrical framework, the gauge fields and their associated ghost and
antighost fields occurring in quantized gauge-affine gravity have been
described through a$\ \overline{GL}(4,%
\mathbb{R}
)\otimes \overline{GA}(4,%
\mathbb{R}
)$-superconnection. Furthermore, in order to realize supercurvature
constraints we introduce over a principal superfiber bundle an even
pseudotensorial 1-superform which permits us to introduce a generalized
superconnection, and by applying the exterior covariant superdifferential
this gives the generalized supercurvature. Then the supercurvature
constraints are determined by the fact that the generalized supercurvature
is an even tensorial 2-superform which leads to the determination of \ the
gauge-affine gravity BRST and anti-BRST transformations. The obtained BRST
transformations are nilpotent and equivalent to those given in \cite%
{Mielke11,neeman22,mez}. Reducing the four-dimensional general affine group
double-covering $\overline{GA}(4,%
\mathbb{R}
)$ to the Poincar\'{e} group double-covering $\overline{ISO}(1,3)$ we have
also found the BRST and anti-BRST transformations of the fields present in
quantum Einstein gravity. Moreover, we have shown how to construct the
gauge-fixing superaction for gauge-affine gravity in analogy with what is
realized in super-Yang-Mills theory \cite{tah1,tah12}. Its lowest component
represents the gauge-fixing action and is invariant under the geometrical
BRST transformations. By using the fact that the dynamic field occurring in
Einstein's gravity is represented by the tensor metric $g_{\mu \nu \text{\ }%
} $and following the same spirit of construction of the superaction as in 
\cite{tah1,tah12} we have found the gauge-fixing action for quantum Einstein
gravity recovering then the standard results \cite{nakanishi}.

\bigskip

{\LARGE Acknowledgments\bigskip }

M.T. acknowledges support from the Alexander von Humboldt Stiftung. He
thanks Prof. W. R\"{u}hl ( Department of Physics, Kaiserslautern University
of Technology) and Prof. H.D. Doebner ( Department of Physics, Metallurgy
and Material Science, Technical University of Clausthal) \ for hospitality.

\end{document}